\documentclass[prl,aps,twocolumn]{revtex4-2}
\usepackage{graphicx} 
\usepackage[usenames]{color}
\usepackage{amsmath,amssymb}
\usepackage{gensymb}
\usepackage{natbib}
\usepackage{xcolor}
\def\strutdepth{\dp\strutbox}
\def\nw#1{\strut\vadjust{\kern-\strutdepth\vtop to0pt{\vss\hbox to\hsize
{\hskip\hsize\hskip5pt$\leftarrow$\hss\strut}}}{\em #1}}
\usepackage{rotating}

\usepackage{color}
\definecolor{red}{rgb}{1,0,0}

\usepackage{blindtext}

\begin{document}

\title{Non-Local Particle Flows Become Local When Considering Dissipative Stress}

\author{Martin Trulsson}
\altaffiliation{Visiting Professor at: Department of Environment, Land and Infrastructure Engineering (DIATI), Politecnico di Torino, Torino, Italy}
\affiliation{Computational Chemistry, Lund University, Sweden}

\begin{abstract}
Dense granular and suspension flows under inhomogeneous shear exhibit persistent particle motion in regions where the local yield criterion is subcritical, an apparent breakdown of locality that has motivated the development of a generation of nonlocal rheological models. Using particle-resolved simulations of frictionless dense suspensions in two-dimensional Kolmogorov flow, we show that two independent considerations together account for this signature.
First, replacing the conventional shear stress by a shear-rate-weighted dissipative stress $\tau_W=\langle \tau \dot \gamma \rangle/\langle \dot \gamma \rangle$, which isolates the component of stress that performs irreversible work, restores the homogeneous $\mu(J)$ law throughout the bulk of the flow, with the inferred friction remaining strictly above yield. Second, a simple geometric mixing-length construction, applied with conventional stresses and requiring no fluctuation input, accounts for the residual sub-yielding within a sub-diameter layer at flow reversals. Each approach is based on a different philosophy and mechanism, and together they suggest that much of the apparent non-locality in this geometry and frictionless case is an artefact of how stress is measured and averaged rather than an intrinsic breakdown of local rheology.
 \end{abstract}

\date{\today}
\maketitle
\begin{figure}[!htbp]
\includegraphics[width=0.5\textwidth]{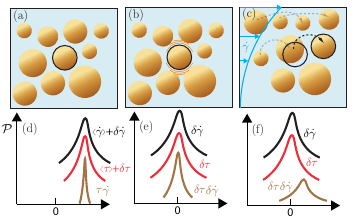}
\caption{Sketch of the processes leading to non-local stresses. (a) A schematic configuration.  (b) A particle oscillates reversibly within its cage. (c) The particle undergoes a finite displacement (strain) and dissipates energy. (d) Sketch of the distributions, $\mathcal{P}$, of strain rate $\dot \gamma(t)$, $\tau(t)$, and $\tau(t) \dot \gamma(t)$. The strain rate and stress can further be decomposed into mean, $\langle \cdot \rangle$, and fluctuating, $\delta$, components.  For homogeneous flows and at high strain rates, fluctuations are negligible compared to the mean. Near shear reversals, however, fluctuations become significant, leading to two possible situations: (e) uncorrelated fluctuations in strain rate and stress, or (f) correlated fluctuations.}
\label{fgr:sketch}
\end{figure}

Dense granular and suspension flows exhibit a transition from liquid‑like to solid‑like behaviour as the packing fraction $\phi$ approaches jamming. Under homogeneous shear, this transition is quantitatively captured by local $\mu(X)$-rheologies, in which the stress ratio $\mu =\tau/P$ is expressed as a function of an inertial number $I=\frac{\langle d\rangle \dot \gamma}{\sqrt{P/\rho}}$~\cite{MiDi04,Cruz05,Jop06}, a viscous number $J=\frac{\eta_f \dot \gamma}{P}$~\cite{Boyer11}, or their visco‑inertial combination $K=J+(St_c)^{-1} I^2$~\cite{Trulsson12}, $St_c$ being the Stokes number for the visco-inertial transition. These frameworks, all embedding a clear separation between flowing ($\mu>\mu_c$) and non-flowing ($\mu<\mu_c$) states, where $\mu_c$ is an effective critical Coulomb friction coefficient, establish a one‑to‑one correspondence between $\mu$ and the local shear‑rate and viscous number. Their success has shaped much of our current understanding of dense particulate matter in steady, homogeneous conditions~\cite{Forterre08,Guazzelli18}.
However, when applied to inhomogeneous flows -- such as pressure‑driven configurations or geometries with strong spatial gradients -- local rheologies systematically fail. Simulations and experiments reveal persistent creeping or sub‑yield motion in regions where $\mu<\mu_c$, in apparent contradiction with the local constitutive relation~\cite{Koval09, Kamrin12, Bouzid13, Bouzid15epl, Gillissen20,Bhowmik24}. To reconcile these observations, several non-local models have been proposed, typically introducing a ``fluidity'' field with a finite relaxation length~\cite{Bocquet09,Pouliquen09,Kamrin12,Henann13} or invoking a rescaled granular temperature $\Theta$ to capture cooperative rearrangements~\cite{DeGiuli17,Trulsson17,Pahtz19,Kim20,Gaume20, Bhowmik24,Bhowmik25}. Although these approaches successfully reproduce non-local flow profiles, they do not identify the microscopic mechanism \cite{Bouzid15} that generates the apparent breakdown of locality,  even if Eyring and kinetic interpretations have been put forward \cite{Zhang17,Berzi24}.
Here we advance a different perspective. We propose that the observed non-locality arises not from a failure of local rheology itself, but from the way stress is defined and averaged in strongly fluctuating, spatially heterogeneous flows. To test this idea, we developed two distinct and logically independent approaches.
The first concerns how stress is measured: here we replace the conventional instantaneous shear stress by a dissipative stress $\tau_W$, defined as the shear-rate-weighted stress
\begin{equation}
\tau_W(y) = \frac{\int \tau(t,y) \dot \gamma(t,y) \, dt}{\int \dot \gamma(t,y) \, dt},
\label{eq:diss}
\end{equation}
where $y$ denotes the coordinate along the gradient direction. This definition reflects the physical fact that energy dissipation only occurs through irreversible deformation; fluctuations of $\dot \gamma$ around zero at fixed $\tau$ -- corresponding to reversible (configurational), non-dissipative responses -- do not contribute (see Fig.~\ref{fgr:sketch} for a schematic of the involved processes). Thus, $\tau_W$ filters out reversible stress fluctuations and isolates the part of the stress that is actually responsible for flow, eliminating the apparent sub-yielding signature throughout the bulk of the flow.
The second concerns how stress is averaged: a simple mixing-length construction, in which the local $\mu(J)$ law is convolved over a particle-sized region, accounts for the residual sub-yielding within a sub-diameter layer at flow reversals, without invoking any fluctuation-based correction. 

\begin{figure}[!htbp]
\includegraphics[width=0.4\textwidth]{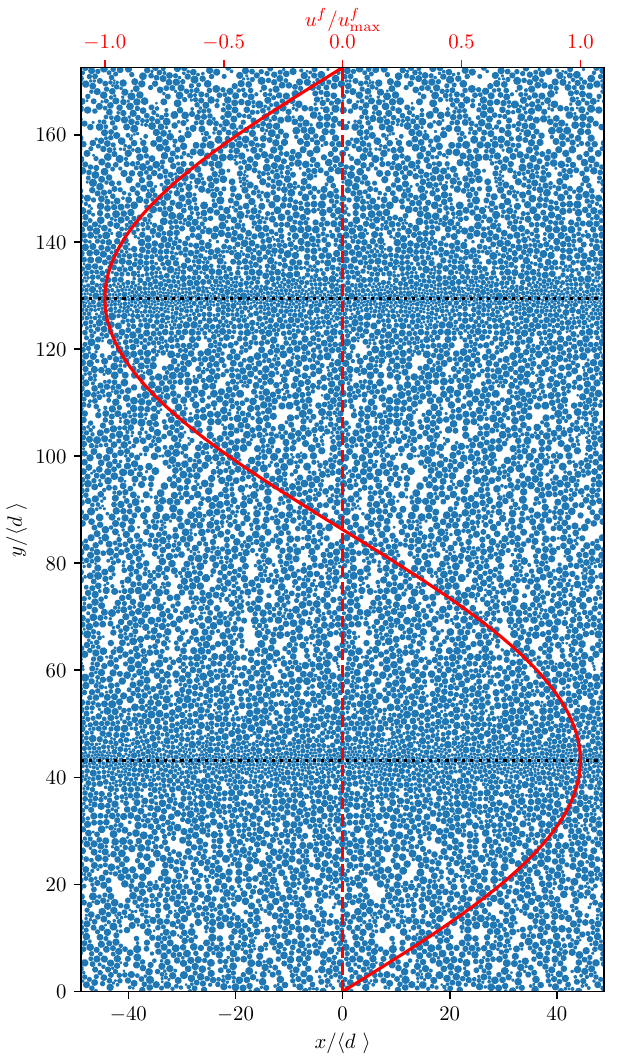}
\caption{Snapshot of the simulation cell showing the system dimensions and the flow field. Size segregation is evident: smaller particles preferentially occupy the low-shear-rate region, whereas larger particles are found predominantly in the high-shear‑rate regions.}
\label{fgr:snapshot}
\end{figure}


\begin{figure*}[!htbp]
\includegraphics[width=0.24\textwidth]{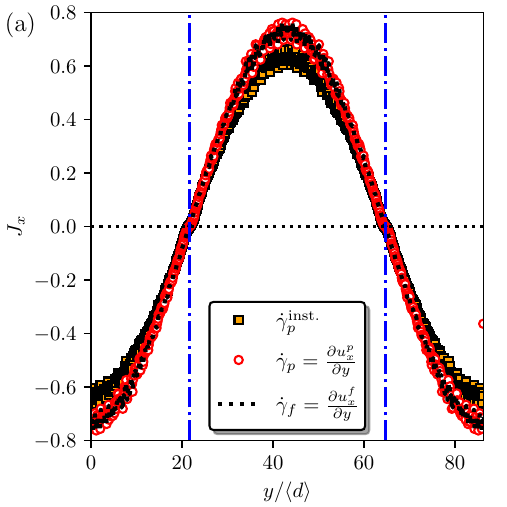}
\includegraphics[width=0.24\textwidth]{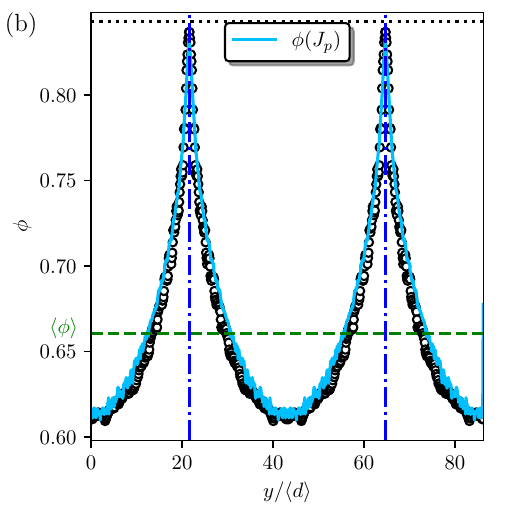}
\includegraphics[width=0.24\textwidth]{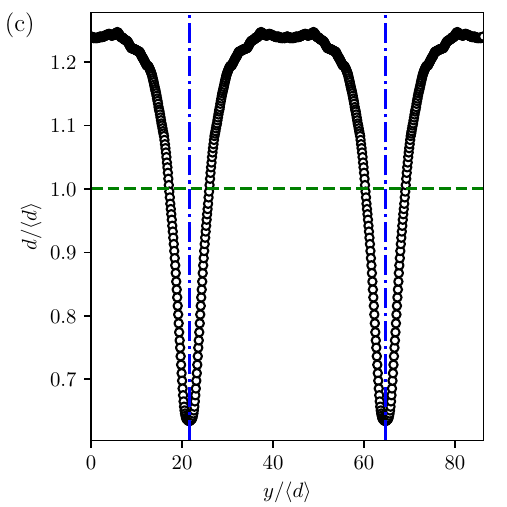}
\includegraphics[width=0.24\textwidth]{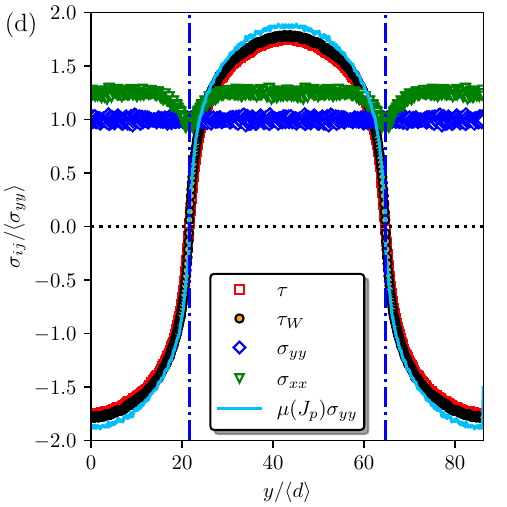}
\includegraphics[width=0.24\textwidth]{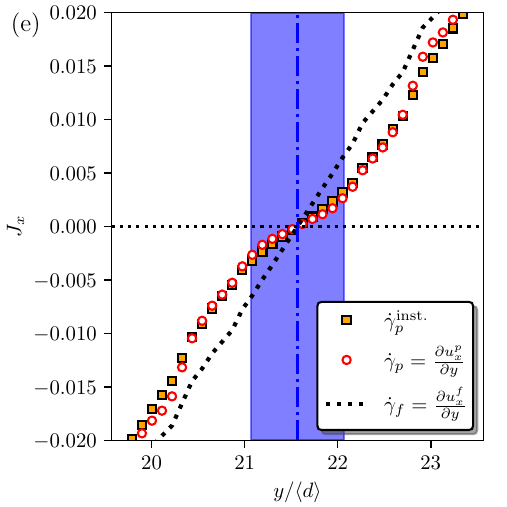}
\includegraphics[width=0.24\textwidth]{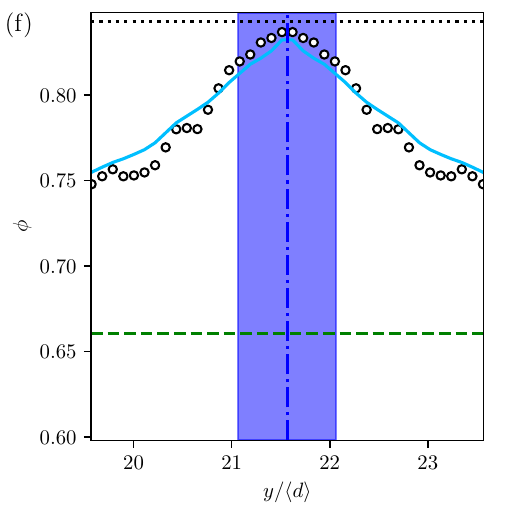}
\includegraphics[width=0.24\textwidth]{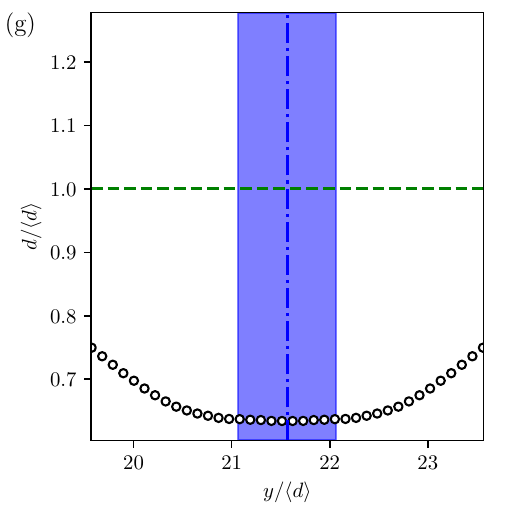}
\includegraphics[width=0.24\textwidth]{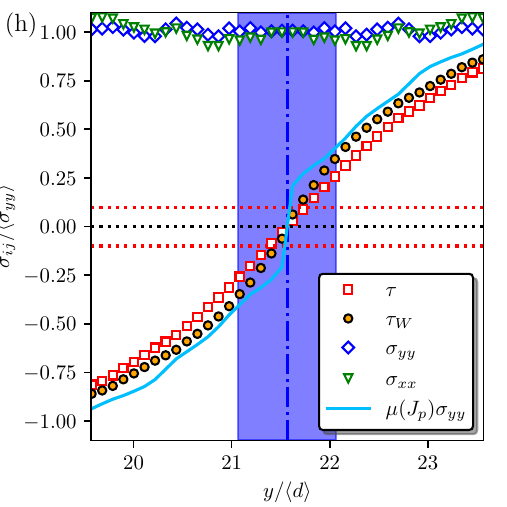}
\caption{Local fields versus position in the Kolmogorov flow: (a) viscous number \(J\), computed using different measures of \(\dot\gamma\); (b) packing fraction \(\phi\); (c) mean particle diameter \(d/\langle d\rangle\); and (d) stress fields, including the normal stress \(\tau\), dissipative stress \(\tau_W\), and normal stress components \(\sigma_{xx}\) and \(\sigma_{yy}\). Panels (e–h) show the corresponding quantities zoomed in near the shear-reversal regions. Turquoise lines in (b) and (e) indicate the prediction for \(\phi\) from the standard local rheology \(\phi(J)\). Turquoise lines in (d) and diamond symbols in (h) show the predicted shear stress from the local rheology \(\mu(J)\), using the measured normal stress \(\sigma_{yy}\).}
\label{fgr:local}
\end{figure*}

\emph{Results -}
Fig.~\ref{fgr:snapshot} shows a snapshot of the simulation cell. We begin by examining spatial variations across an $\sim 86\langle d\rangle$-wide Kolmogorov flow (see End Matter for simulation details, statistical averaging and Gaussian filtering) and compare them to results predicted from steady state (SS) results of the same system. The viscous number $J(y)$ (see Figs.~\ref{fgr:local}(a,e)), which reflects the local shear rate, follows the imposed sinusoidal velocity field ($u_{x}^f = u_{\rm max}^f \sin(2\pi y/L_y)$), but exhibits pronounced deviations near the reversal points where $J$ changes sign. 
Local instantaneous shear-rate estimates using Gaussian filtering and velocity fields obtained from particle-averaged motion yield nearly identical $J$-profiles, particularly close to shear reversals, confirming the robustness of the Gaussian estimate.
Turning to the packing fraction, the expected trends from the local $\phi(J)$-rheology are clearly recovered: the density peaks at the minima of $|J|$, \emph{i.e.,} near the reversals (see Figs.~\ref{fgr:local}(b,f)). These regions reach values of up to 84\%, consistent with random close packing for discs. The strong inhomogeneity in packing is accompanied by a pronounced size segregation: small particles accumulate in the low-$|J|$ regions, while larger particles migrate toward the high-shear zones. This pattern mirrors the classical Brazil‑nut–type segregation familiar from granular flows~\cite{Gray18}. Both simulation snapshots and average particle‑diameter profiles (see Figs.~\ref{fgr:snapshot} and \ref{fgr:local}(c,g)) show this effect unambiguously.

We now examine the stress fields in Figs.~\ref{fgr:local}(d,h). The normal stress perpendicular to the flow, $\sigma_{yy}$, remains spatially uniform, while $\sigma_{xx}$ is nearly uniform with a characteristic dip near the flow reversals, consistent with previous observations \cite{Bhowmik24}. The clear normal stress difference observed in these systems remains unexplained \footnote{Substituting $\sigma_{xx}$ or $P = (\sigma_{xx} + \sigma_{yy})/2$ into the $\tau$ or $\tau_W$ leaves the qualitative results unchanged.}.
A striking contrast emerges when comparing the two definitions of shear stress. With the conventional instantaneous stress, $|\tau|/\sigma_{yy}$ crosses zero whenever the shear rate changes sign, producing extended regions where the inferred stress ratio falls well below the yield value (equivalently, $\mu<\mu_c$). In contrast, the dissipative stress $|\tau_W|/\sigma_{yy}$ remains close to or above $\mu_c$ throughout most of the domain \footnote{Deviations between the measured stress and the local prediction at high $|J|$ are merely a consequence of the limited accuracy of our fits of the SS data in the high-$J$ region, where the simple fit slightly overestimates \emph{e.g.,} the effective friction coefficient.}. 
Only within a narrow layer, typically less than one particle diameter from the reversal, does $|\tau_W|/\sigma_{yy}$ dip slightly below $\mu_c$. These points should be interpreted with caution, as we have limited accumulated strain in this region (see Appendix, Fig.~\ref{fgr:strain}), and excluding all data within $0.5\langle d\rangle$ of the reversal removes these deviations entirely.

\begin{figure}[!htbp]
\includegraphics[width=0.23\textwidth]{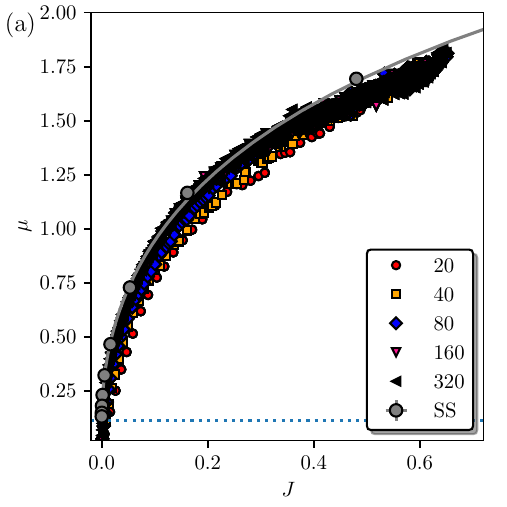}
\includegraphics[width=0.23\textwidth]{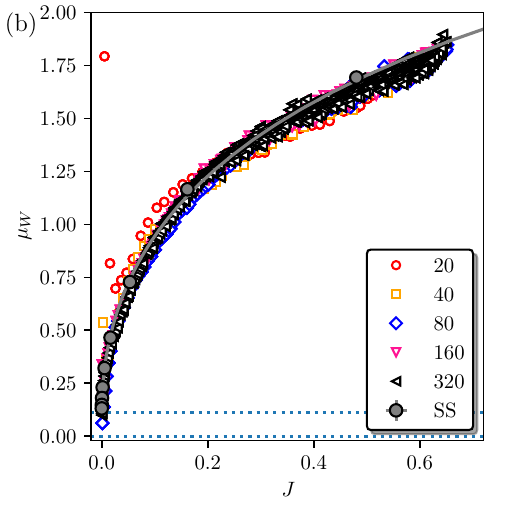}
\includegraphics[width=0.23\textwidth]{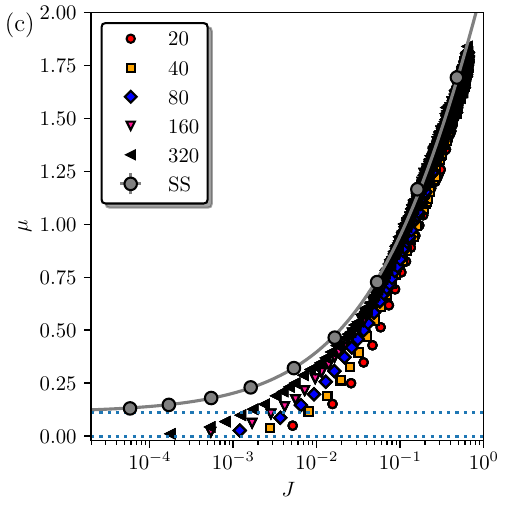}
\includegraphics[width=0.23\textwidth]{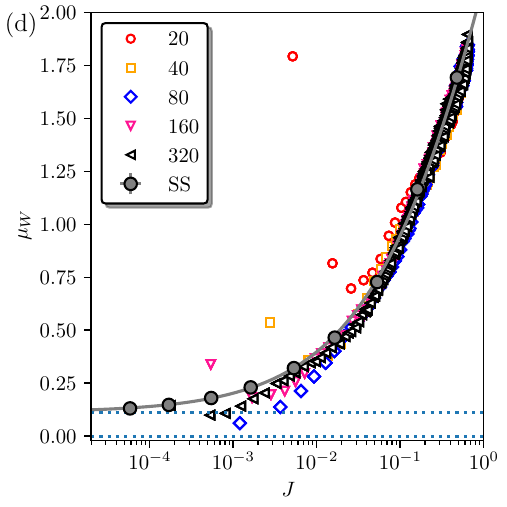}
\caption{
Comparison of effective friction coefficients versus viscous number $J$: (a,b) normal friction coefficient $\mu$, and (c,d) dissipative friction coefficient $\mu_W$. Data are shown in both lin–lin (a,c) and lin–log (b,d) representations. SS denotes the steady-state result obtained under homogeneous shear-rate conditions. The legend indicates the system height, $L_y/\langle d\rangle$. The grey solid line corresponds to the best fit to the SS data (see End Matter for details).}
\label{fgr:MuJ}
\end{figure}

Next, we evaluate the effective friction as a function of $J$ across the full system, for several system sizes, comparing both stress definitions with homogeneous steady-shear (SS) data.
In both linear-linear and linear-log representations, the standard stress measure produces an apparent decrease of $\mu$ below the yield friction at small $J$. In general, the deviation from the local rheology is larger for smaller system sizes (see Fig.~\ref{fgr:MuJ}).
Using the dissipative stress, however, all data collapse onto a single master curve that coincides with the SS results. This collapse holds over the entire accessible $J$-range, with only minor noise at the lowest $J$ values for each system size. These lowest points correspond to the narrow shear‑reversal layer, where the accumulated strain is small and statistical fluctuations are naturally larger. In the smallest systems, this manifests as a slight apparent increase in the friction, but the overall trend remains clear: when stress is measured through dissipation per strain, the friction law is only weakly influenced by the strong inhomogeneity of the Kolmogorov flow. This is despite the large particle segregation occurring in these systems.  
Most importantly, as $J\to 0$, the measured friction remains above the yield value $\mu_c$ obtained from SS. This resolves the long‑standing puzzle of apparent sub‑yielding in non‑local flows: there is no sub‑yielding when the stress is defined through dissipation, which effectively filters out the reversible elastic components. Finally, the packing‑fraction data closely follow the local $\phi(J)$‑rheology, consistent with earlier findings. 

\begin{figure}[!htbp]
\includegraphics[width=0.23\textwidth]{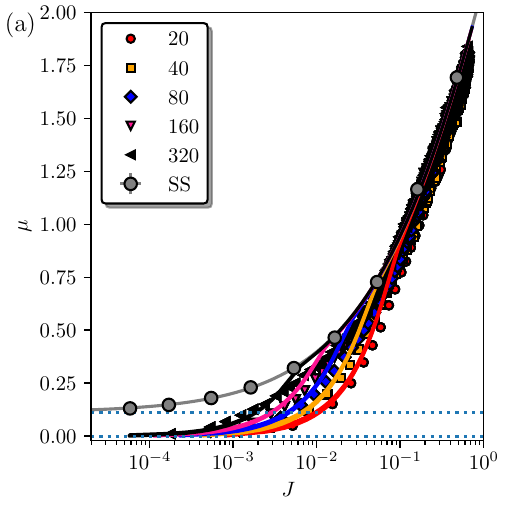}
\includegraphics[width=0.23\textwidth]{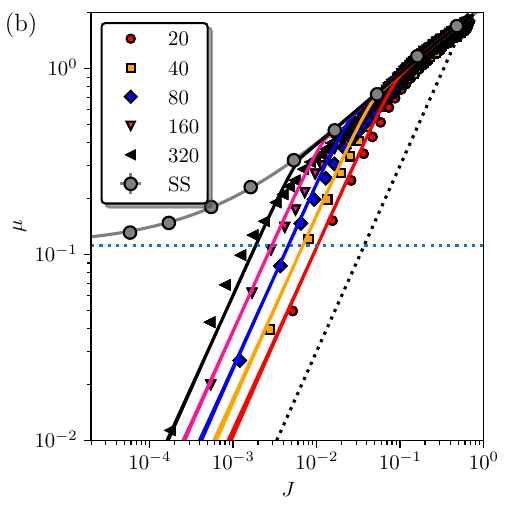}
\caption{Local rheology using a mixing-length of size $d$. Colours as in previous figures. This simple model captures the sub-yielding.
(a) Lin-log and (b) log-log plot. The black dashed line indicates a scaling $\mu \sim J$.}
\label{fgr:MuFluid}
\end{figure}

Particles located far from the reversal points ($J=0$) are surrounded by shear flows of a single sign, whereas those near a reversal are surrounded by shear flows of opposite sign. A particle straddling this boundary therefore samples two competing deformation modes: both the shear rate and shear stress contributions change sign across its extent, leading to partial cancellation when stresses are averaged over a particle‑sized region. To model this effect, we apply the local $\mu(J)$ rheology but incorporate a mixing length of one grain diameter $d$, similar to that proposed by Nott and Dsouza \cite{Nott17, Nott20}. We treat a particle of diameter $d$ whose centre is positioned at varying distances from the reversal surface, and compute a mixed stress $\mu^{\rm mix}(J)$ by segment‑averaging the local stress predicted by the $\mu(J)$-law using the measured local $J_p$ and $\sigma_{yy}$.
This simple construction captures the sub‑yield region with surprising accuracy (see Fig.~\ref{fgr:MuFluid}) and indicates that sub‑yielding is confined to a layer thinner than a single particle diameter.
It is worth emphasising that the mixing-length argument is logically independent of the dissipative-stress analysis presented above. The construction uses only the conventional local $\mu(J)$ law and the geometric fact that a particle straddling a reversal samples shear of opposite sign across its diameter; it makes no reference to fluctuation correlations, granular temperature, or $\tau_W.$ That this purely geometric model reproduces the sub-yield region with a thickness of order $\langle d\rangle$ -- and predicts the emergent linear scaling $\mu \sim J$ observed both here (see Fig.~\ref{fgr:MuFluid}(b)) and in \cite{Gillissen20} -- indicates that even the residual non-locality near reversals can be attributed to particle-scale averaging rather than to a breakdown of local rheology. The two analyses thus converge on the same conclusion through different routes: $\tau_W$ restores locality in the bulk by filtering reversible fluctuations, while the mixing-length argument explains the residual sub-yielding near reversals as a discretisation artefact of the continuum limit due to the particle discreteness.
This further justifies excluding sub-diameter zones near reversals when comparing to local rheology.
\emph{Discussion -}
We have presented two independent arguments that together account for the apparent non-locality observed in dense suspensions under Kolmogorov flow. The dissipative stress $\tau_W$ obeys the homogeneous $\mu(J)$ law throughout the bulk of the flow, and a geometric mixing-length construction accounts for the residual sub-yielding within a sub-diameter layer at reversals; neither argument requires the apparent non-locality to be intrinsic.
The constitutive relation between dissipative friction and the viscous number holds independently of the degree of flow inhomogeneity, indicating a simple underlying mechanism: the energy dissipated per unit strain remains invariant across flow conditions. This observation provides a physical explanation for why previous non‑local models based on power‑law scalings of stress and granular pressure successfully collapsed data, while offering little insight into why such collapses work.

The dissipative formulation clarifies the role of fluctuations. As the average stress and shear rate tend to zero, their fluctuations and, crucially, their correlations become increasingly important (see Fig.~\ref{fgr:sketch}). 
When stress and shear-rate fluctuations are uncorrelated, $\langle \tau \dot \gamma \rangle=\langle \tau \rangle \langle \dot \gamma \rangle \to 0$ and the inferred dissipation vanishes; when they are tightly correlated through an effective viscous response, $\langle \tau \dot \gamma \rangle$ remains finite and encodes the granular temperature implicitly (see Appendix).
In this sense, the dissipative stress formalism embeds the effect of granular temperature without requiring its explicit measurement: the correlation structure is already encoded in $\langle \tau\,\dot\gamma\rangle$. 
We test this connection directly in the Appendix, finding that $(\mu_W-\mu) J$ scales approximately as $\Theta_J^2$ across all system sizes.
This framework thus bypasses the need for granular temperature as an independent state variable and offers a more transparent route to predicting dense granular and suspension flows. While the formulation does not directly yield the stress field, this may be of secondary importance, since stress enters primarily through its role in determining energy dissipation and, ultimately, flow resistance. Remarkably, the approach remains robust even in the presence of pronounced particle‑size segregation, as long as the microstructure remains amorphous. 
A complementary kinetic-theory perspective is developed in Ref.~\cite{Alessio26}, where non-local granular fluidity is derived perturbatively from a granular temperature equation. Our framework reaches the same physics from the opposite direction, absorbing the relevant fluctuation correlations directly into $\langle \tau \dot \gamma \rangle$ without postulating $\Theta_J$ as an independent state variable.

These findings raise an intriguing question: are these results particular to frictionless dense suspensions in Kolmogorov flows, or does the same dissipative formalism extend to frictional grains or non-isotropic shaped particles \cite{Trulsson18,Farnaz22,Bhowmik25}, soft or cohesive particles \cite{Bouzid15, Coulomb17, Mandal21}, or more general flow geometries \cite{Koval09, Kim20}? Exploring this possibility may offer a unified perspective on locality in dense disordered materials.\\

\bibliography{Reversal}

\onecolumngrid
\vspace{1em}  
\begin{center}
    \textbf{\large End Matter}
\end{center}
\vspace{1em}
\twocolumngrid

\subsection{DEM simulations details}
\label{Sim}
We use the discrete element method to simulate particles in Kolmogorov fluid flows, where $u_{f,x}(y) \sim \sin(2\pi y /L)$ and zero in $u_{f,y}$. We keep the length in the x-direction fixed at $L_x/\langle d \rangle\simeq48$, while we vary the height of the system $L_y/\langle d \rangle$ between 22 and 345. We keep a fixed average packing fraction $\langle \phi \rangle=0.66$, far away from the critical shear jamming packing fraction at $\phi_J=0.84$. Hence, the number of particles ranged from approximately 800 to 12800. 
Particles interact via simple harmonic interactions where the normal force is proportional to the particle normal overlap $\vec{\delta}_n$ as  $\vec{f}_n = k_n \vec{\delta}_n$, where $k_n$ is the normal spring constant. 
In addition to contact interactions, the particles experience Stokes drag and torque, $\vec{f}_{\rm Stokes}= -3 \pi \eta_f(\vec{u}_p-\vec{u}_f(y))$ and $\vec{\tau}_{\rm Stokes}= -\pi \eta_f d^2 (\omega_p-\omega_f(y))$, where $\omega_f(y)=0.5\frac{\partial u_{f,x}}{\partial y}$ and $\eta_f$ is the fluid viscosity. The particles also experience short-range lubrication forces, $\vec{f}_{\rm sq.} \propto \frac{\eta_f d_{ij}}{h_{ij}+\epsilon} (\Delta \vec{u}_{ij} \cdot \vec{n}_{ij}) \vec{n}_{ij}$ and $\vec{f}_{\rm sh.} \propto \eta_f \log(\frac{d_{ij}}{h_{ij}+\epsilon}) (\Delta \vec{u}_{ij}\cdot \vec{t}_{ij}) \vec{t}_{ij}$, where $d_{ij}= \frac{2 d_i d_j}{d_i+d_j}$ is the reduced diameter, $h_{ij}$ the closest gap between the particles, $\vec{n}$ and $\vec{t}$ are the normalised normal and tangential directions respectively, and $\Delta \vec{u}_{ij}$ the relative velocities between particles $i$ and $j$ evaluated at the closest contact points (i.e., including both translational and rotational contributions). These lubrication forces are shifted and truncated at $h_{\rm cut}$, beyond which the lubrication forces are zero. If particles are overlapping, \emph{i.e.} $h_{ij}<0$, we choose to saturate the distance-dependent term of the lubrication setting $h_{ij}=0$.
Particles have a mass density of $\rho$ and are assumed to be isodense with the fluid. The corresponding maximum Stokes number is kept $St^{\rm max}= \rho \dot \gamma^{\rm max}/\eta_f<0.1$ and the softness parameter $So=P/k_n > 10^3$. We furthermore use a polydisperse system where the particle diameters are uniformly distributed between $[0.5\langle d \rangle, 1.5 \langle d \rangle]$. 
When sampling instantaneous stresses and shear-rates, we apply a Gaussian filter between each pair of particles as $W_{ij} \propto e^{-\alpha (h_{ij}-h_{\rm cut})^2}$ if $(h_{ij}-h_{\rm cut})>0$, one otherwise.
We chose a filtering size of $\alpha=(10/\langle d \rangle)^2$. Instantaneous shear-rates on the individual particle were then $\dot \gamma_{i}= \sum_{{\rm neighbours}\, j} \frac{W_{ij} \Delta u_x \Delta y}{\Delta y^2}/\sum_{{\rm neighbours}\, j} W_{ij}$, similar as proposed in Ref.~\cite{Bouzid15}. In a similar way, the instantaneous stress energies (stress times an area)  were calculated as 
$\hat{\sigma}_{kl,i} = \sum_{{\rm neighbours}\, j} f_{ij,k} r_{ij,l}$, with $\hat{\tau}_{i}=\hat{\sigma}_{yx,i}=\hat{\sigma}_{xy,i}$. The numerator in Eq.~\ref{eq:diss}  was evaluated on the particle scale as
$\frac{1}{A}\sum_{i\in  A} \dot \gamma_{p,i} \hat{\tau}_{p,i} \frac{4s_{i}}{\pi d_i^2}$, where $A$ is the sampling area, usually the bin area, and $s_i$ is the area of disc overlapping with the sampling volume.
Since we bin our data in the y-direction, this corresponds to segments of the discs. Similarly, we do with $\dot \gamma_p = \sum_{i \in A} \dot \gamma_{i} s_{i}/\sum_{i \in A} s_{i}$.
Similarly, the packing fraction $\phi = \frac{1}{A}\sum_{i \in A} s_{i}$ , conventional stresses where sampled $\sigma_{kl} = \frac{1}{A} \sum_{i \in A} \hat{\sigma}_{kl,i} \frac{4s_{i}}{\pi d_i^2}$, velocities $u_{p,x}=\sum_{i \in A } u_{x,i} s_{i}/\sum_{i \in A }s_{i}$, and average diameter $d = \sum_{i \in A} d_i \frac{4 s_{i}}{\pi d_i^2}/\sum_{i \in A} \frac{4 s_{i}}{\pi d_i^2}$.
Note that many quantities can be defined per particle, per total area, or per solid or fluid area. All quantities are time-averaged after a first initial transient (including particle migration of big and small discs). Notice that the particle scale dissipation could also be evaluated through true dissipation on the particle scale, but since the denominator also includes $\dot \gamma_p$, we decided to stick with this formulation, which also makes the connection to previous dissipative measures simpler. We chose a $h_{\rm cut}/\langle d \rangle=0.05$ and $\epsilon/\langle d \rangle=0.001$. We bin our data in sizes of $\Delta y \approx 0.1 \langle d \rangle$. All forces involving the fluid viscosity (drag, torque, and lubrication) are systematically divided by $d_{\rm eff}$. This preserves the dimensions of $\eta_f$ as in 3D, so that $J$ requires no correction.  

\newpage
\onecolumngrid
\vspace{1em}  
\begin{center}
    \textbf{\large Appendix}
\end{center}
\vspace{1em}
\twocolumngrid

\subsection{Rheology at homogeneous shear rate fields}
In the case of linear fluid velocity cases, one obtains the classical local $\mu(J)$-rheology. Packing fractions and effective friction coefficients are fitted using the following functional forms: $\phi(J) = \phi_c- a J^\alpha$ (see Fig.~\ref{fgr:phi_mzz}), and $\mu(J) = \mu_c + \frac{b}{(b_0+J^{-\beta})}$ (the form commonly adopted in granular rheology \cite{Jop06}). For our system, we find $\phi_c=0.843 \pm 0.002$, $a=0.266 \pm 0.008$, $\alpha=0.48 \pm 0.02$, $\mu_c=0.112 \pm 0.002$, $b=3.2 \pm 0.1$, $b_0=0.58 \pm 0.06$, and $\beta=0.51 \pm 0.01$. Velocity fluctuations are fitted as $\Theta_J(J)=g J^\gamma$ (see Fig.~\ref{fgr:phi_mzz}), with $g=0.43$ and $\gamma=0.7$ where $\Theta_J=\frac{\eta_f \sqrt{\langle  \delta v_y^2 \rangle }}{\sigma_{yy}\langle d \rangle}$.

\subsection{Rheology removing data points close to the shear reversal points}
Below shows the corresponding $\mu(J)$-rheology after the shear-reversal zones (with width $\langle d \rangle$) have been removed.
Even with this filtering, the standard $\mu$ is clearly lower than the corresponding local rheology, while $\mu_W$ shows an excellent agreement with the local formulation in this span.

 \begin{figure}[!htbp]
\includegraphics[width=0.23\textwidth]{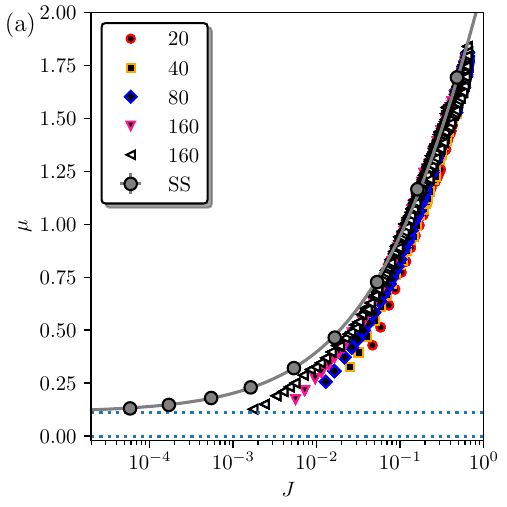}
\includegraphics[width=0.23\textwidth]{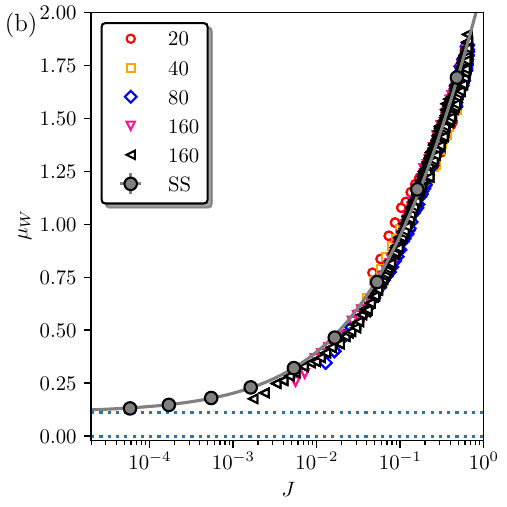}
\caption{As in Fig.~\ref{fgr:MuJ}(c,d) but where the data points in the zones of shear reversals have been removed.}
\label{fgr:MuJclean}
\end{figure}

  \subsection{Varying the Gaussian filter or system size}
\begin{figure}[!htbp]
 \includegraphics[width=0.23\textwidth]{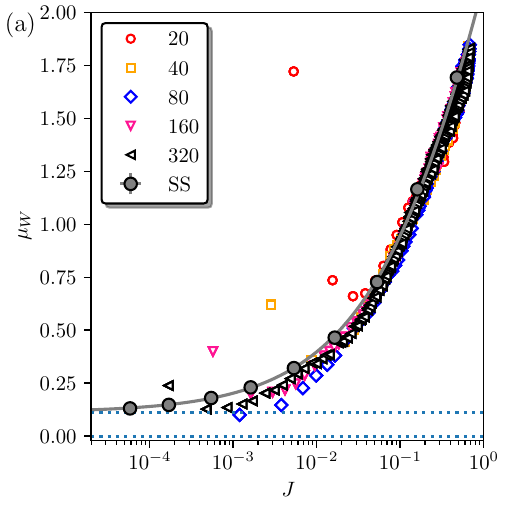}
  \includegraphics[width=0.23\textwidth]{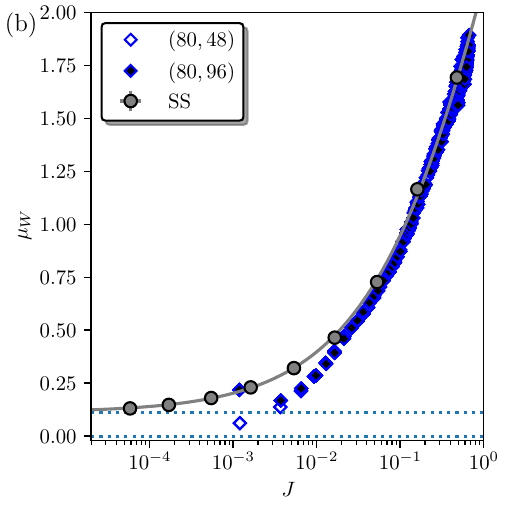}
 \caption{As in Fig.~\ref{fgr:MuJ}(d), but with (a) a $\alpha = (5/\langle d \rangle)^2$ Gaussian filter or (b) varying the system size at fixed $\frac{L_y}{\langle d \rangle}=80$, second number in the legend indicates the width $\frac{L_x}{\langle d \rangle}$.}
\label{fgr:gauss}
\end{figure}

 Varying the Gaussian filter or the system size yields consistent results (Fig.~\ref{fgr:gauss}), as in the main text (see Fig.~\ref{fgr:MuJ}),  confirming that our findings are robust to these choices.
 
 \subsection{Testing the local formulation for an alternative Kolmogorov flow}
\begin{figure*}[!htbp]
 \includegraphics[width=0.23\textwidth]{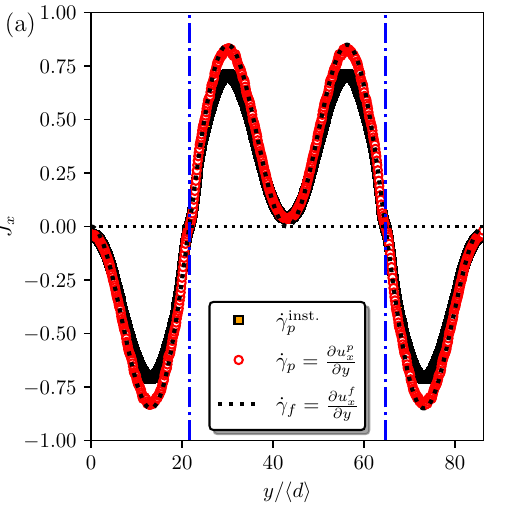}
  \includegraphics[width=0.23\textwidth]{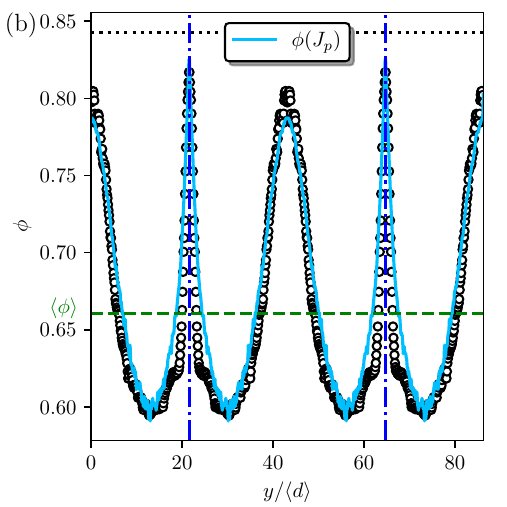}
    \includegraphics[width=0.23\textwidth]{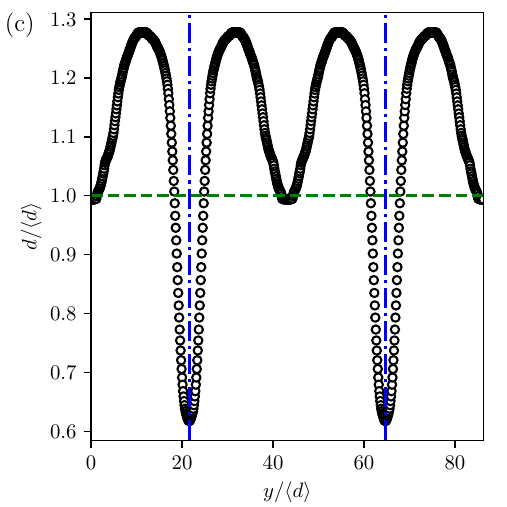}
\includegraphics[width=0.23\textwidth]{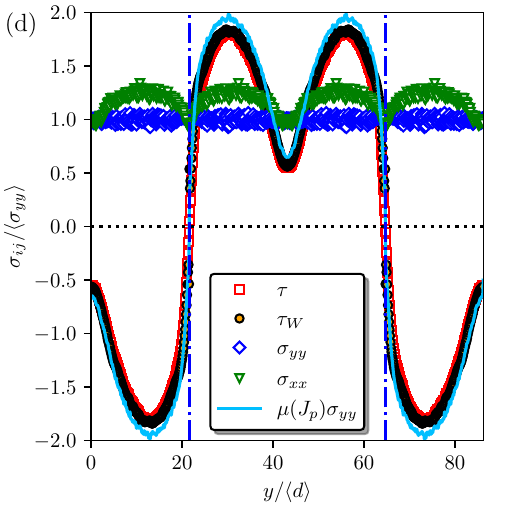}
  \includegraphics[width=0.23\textwidth]{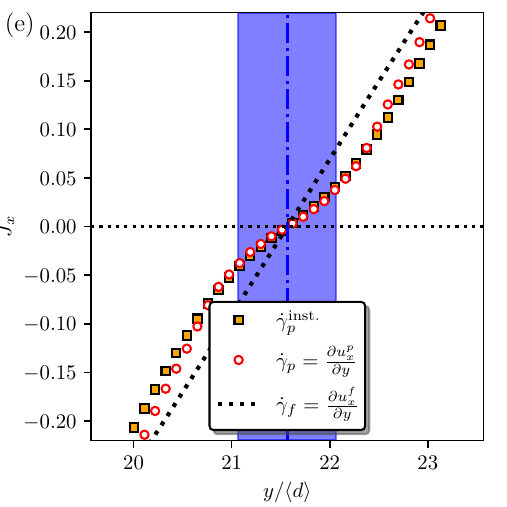}
    \includegraphics[width=0.23\textwidth]{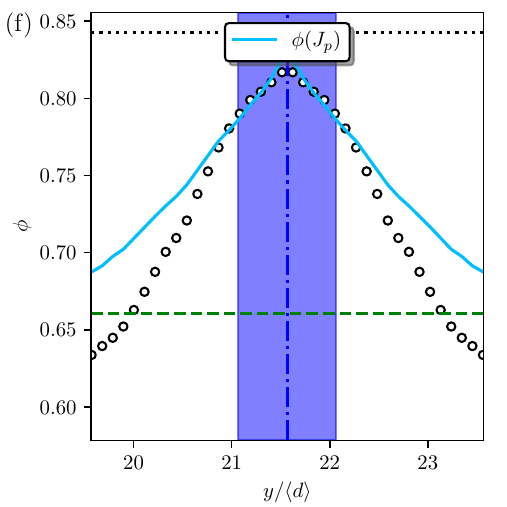}
        \includegraphics[width=0.23\textwidth]{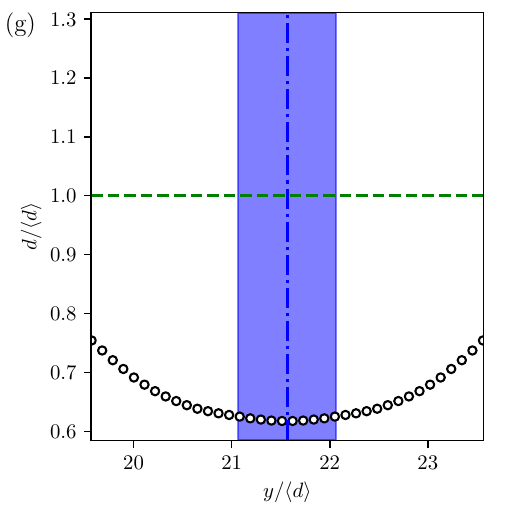}
\includegraphics[width=0.23\textwidth]{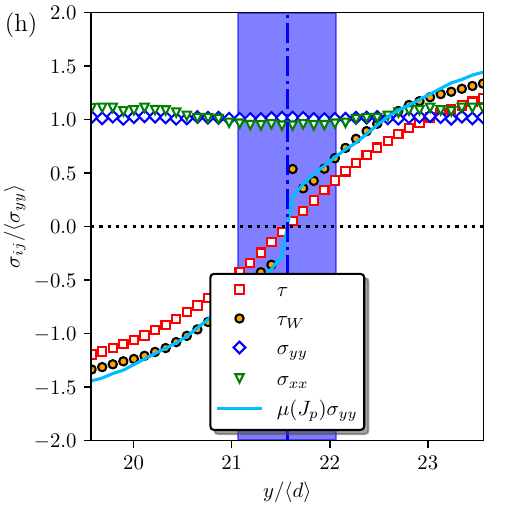}
\caption{As in Fig.~\ref{fgr:local} for a distorted Kolmogrov $u_x^f = u_{\rm max}^f \sin^3(2\pi y/L_y)$ flow.}
\label{fgr:Sin3}
\end{figure*}

 We also apply a distorted Kolmogorov flow as $u_{x}^f = u_{\rm max}^f \sin^3(2\pi y/L_y)$ in Fig.~\ref{fgr:Sin3}, similar to what was studied in \cite{Bhowmik24}.
 We first see that the local rheology fairly well reproduces the $\phi(y)$ and $\mu(y)$-profiles, except close to the reversal zone. 
 Interestingly, the $\phi(y)$ profile is slightly worse than for the normal Kolmogorov flow. We currently lack a clear explanation for this.
 
\subsection{$\phi(J)$ and $\Theta_J(J)$ and local velocity fluctuations}

\begin{figure}[!htbp]
 \includegraphics[width=0.23\textwidth]{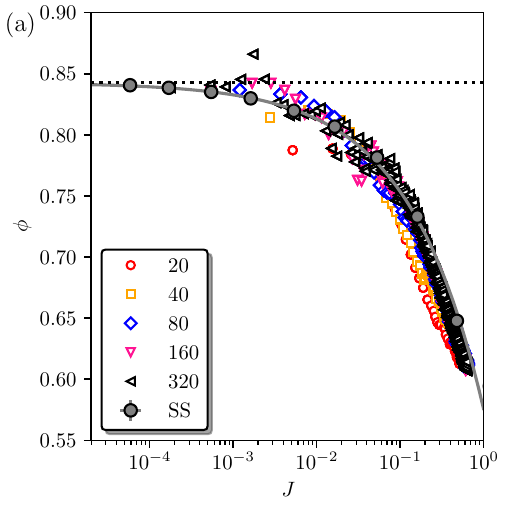}
  \includegraphics[width=0.23\textwidth]{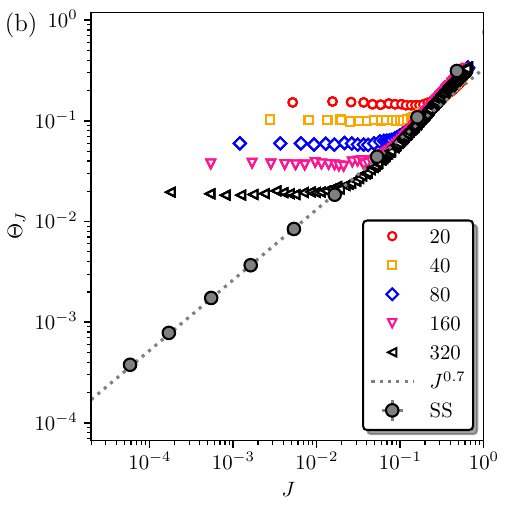}
 \caption{(a) $\phi$ and (b) $\Theta_J$ versus $J$.}
\label{fgr:phi_mzz}
\end{figure}

Fig.~\ref{fgr:phi_mzz} shows both homogeneous $\phi(J)$- and $\Theta_J(J)$-relationships as well as the results for Kolmogorov flow at various flow inhomogeneities (as characterised by the system height $L_y/\langle d \rangle$). 

\begin{figure}[!htbp]
 \includegraphics[width=0.23\textwidth]{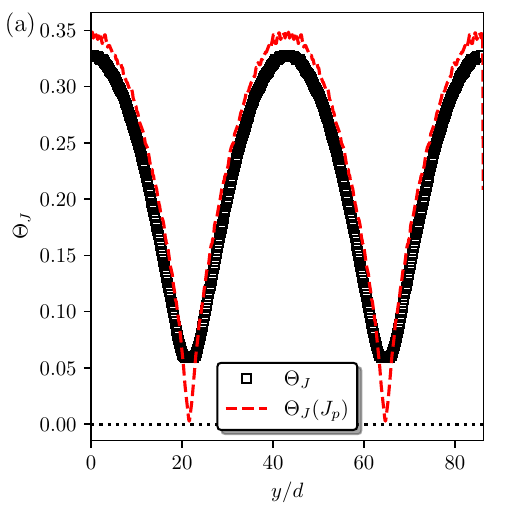}
  \includegraphics[width=0.23\textwidth]{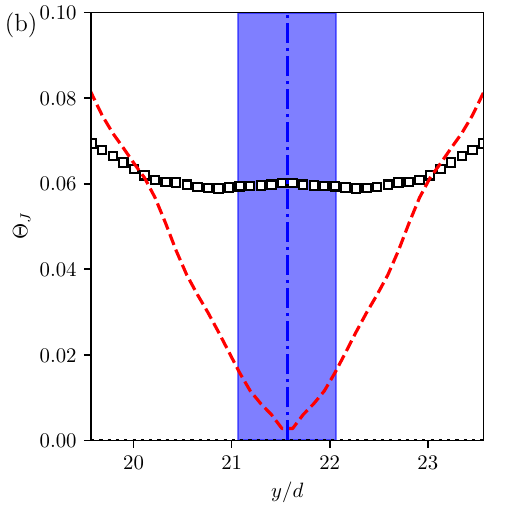}
 \caption{Local rescaled granular temperature $\Theta_J$. (a) Whole range. (b) Shear reversal zone.}
\label{fgr:thetalocal}
\end{figure}

Fig.~\ref{fgr:thetalocal} shows the local $\Theta_J$ as function of $y$-position. Homogeneous works fairly, except in the narrow zones close to the shear reversals.

\subsection{Collapse using $\mu$ and $\Theta$}

\begin{figure}[!htbp]
 \includegraphics[width=0.46\textwidth]{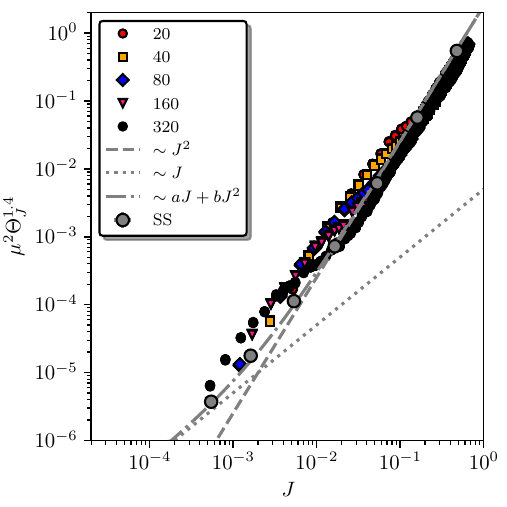}
 \caption{Collapse of the rheology data using $\mu^2 \Theta_J^{1.4}$ versus $J$. The exponents are obtained from steady-state (SS) scalings, where $\mu \sim J^{0.5}$ and $\Theta_J \propto J^{0.7}$, which yield $\mu^2 \Theta_J^{1.4} \propto J^2$ at large $J$ and $\mu^2 \Theta_J^{1.4} \propto J$ at low $J$ (where $\mu \to  \mu_c$), shown as the dashed guide lines together with a cross-over scaling. Symbols and colours as in the main manuscript.}
\label{fgr:collapse}
\end{figure}

Here, we follow previous approaches that collapse the rheology onto a single master curve using both $\mu$ and $\Theta_J$ versus $J$.
Fig.~\ref{fgr:collapse} shows that such an approach is decent but much worse than the previously reported results. We do not yet know why our collapse is worse: whether it is due to the large-scale segregation in our system, to the shear-reversal zone (which is absent in some previous studies), or to a worse fine-tuning of the exponents.

  \section{Relation to granular-temperature-based non-local theories}
\label{app:granular_temperature}
\begin{figure}[!htbp]
 \includegraphics[width=0.23\textwidth]{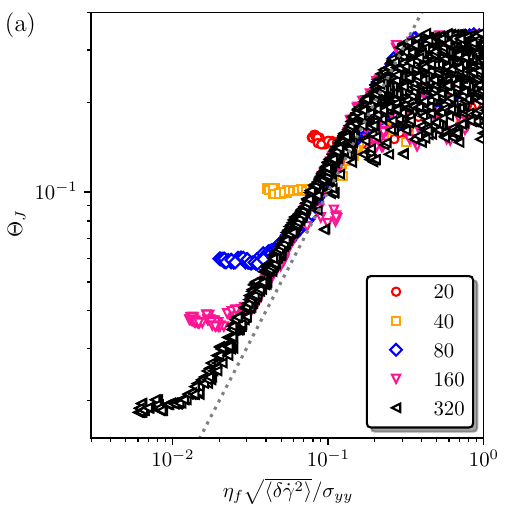}
  \includegraphics[width=0.23\textwidth]{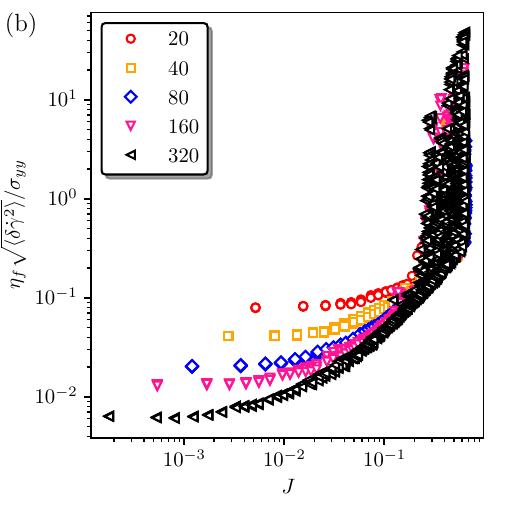}
\caption{(a) Direct test of the relationship $\sqrt{\langle \delta\dot\gamma^2 \rangle} \propto \sqrt{\langle \delta v_y^2 \rangle}/\langle d \rangle$ underlying Eq.~(\ref{eq:gammadot_theta}): $\eta_f \sqrt{\langle \delta\dot\gamma^2 \rangle}/\sigma_{yy}$ versus $\Theta_J$ across all system sizes. The dashed line indicates a one-to-one correspondence, which holds well over the bulk of the data. Deviations appear at the largest values of $\sqrt{\langle \delta\dot\gamma^2 \rangle}$ --- corresponding to the high-$J$, low-density regions at the antinodes of the flow, where the local microstructure becomes sparse, and the assumed local estimate of shear rate starts to fail (for the given Gaussian filtering). (b) The same quantity $\eta_f \sqrt{\langle \delta\dot\gamma^2 \rangle}/\sigma_{yy}$ versus $J$, providing an alternative definition of granular temperature based directly on shear-rate fluctuations rather than velocity fluctuations; the two definitions agree well across the bulk of the $J$ range.}
\label{fgr:shear2}
\end{figure}

The dissipative stress introduced in Eq.~(\ref{eq:diss}) admits a
transparent decomposition that clarifies its relationship to existing
non-local rheologies built on granular temperature. Writing the
instantaneous fields as a mean plus a fluctuation,
\begin{equation}
\tau(t) = \langle \tau \rangle + \delta\tau(t),
\qquad
\dot\gamma(t) = \langle \dot\gamma \rangle + \delta\dot\gamma(t),
\label{eq:decomposition}
\end{equation}
the dissipative stress becomes
\begin{equation}
\mu_W = \frac{\tau_W}{\sigma_{yy}}
= \frac{\langle \tau\,\dot\gamma \rangle}{\sigma_{yy} \langle \dot\gamma \rangle}
= \langle \mu \rangle
+ \frac{\langle \delta\tau\,\delta\dot\gamma \rangle}{\sigma_{yy}\langle \dot\gamma \rangle}.
\label{eq:tauW_decomposition}
\end{equation}
The dissipative stress thus equals the mean stress plus a correction
set by the covariance of stress and shear-rate fluctuations,
normalised by the mean shear rate. In homogeneous steady shear at
moderate $J$, $\langle \dot\gamma \rangle$ greatly exceeds its
fluctuations and $\tau_W \approx \langle \tau \rangle$, so the local
$\mu(J)$ law is recovered identically by both stress measures. Near a
flow reversal, by contrast, $\langle \dot\gamma \rangle \to 0$ while
$\delta\dot\gamma$ remains finite -- particles continue to rattle
within their cages -- so the covariance term dominates and the two
stress measures diverge.
The covariance term in Eq.~(\ref{eq:tauW_decomposition}) is what links
$\tau_W$ to granular-temperature-based non-local
rheologies~\cite{Bhowmik24,DeGiuli17,Trulsson17,Pahtz19,Kim20,Gaume20,Bhowmik25,Zhang17}.
Shear-rate fluctuations are related to velocity fluctuations through
$\langle \delta\dot\gamma^2 \rangle \sim \langle \delta v_y^2
\rangle/\langle d \rangle^2$, a relation we verify directly in Fig.~\ref{fgr:shear2}. This implies that
\begin{equation}
\langle \delta\dot\gamma^2 \rangle \propto \frac{\sigma_{yy}^2}{\eta_f^2}\,\Theta_J^2
\label{eq:gammadot_theta}
\end{equation}
in suspensions. Note that with the convention $\Theta_J \propto \sqrt{\delta v_y^2}$ adopted here (following \cite{Bhowmik24,Bhowmik25}), the correction scales as $\Theta_J^2/J$; under the alternative convention $\Theta \propto \delta v_y^2$ used for granular flows \cite{Zhang17,Kim20}, it is linear in $\Theta$.

\begin{figure}[!htbp]
 \includegraphics[width=0.46\textwidth]{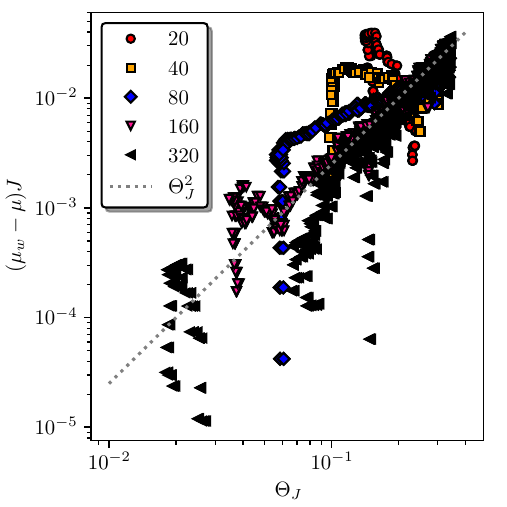}
 \caption{Direct test of Eq.~\ref{eq:covariance_theta}: the rescaled correction ($\mu_W-\mu)J$ plotted against $\Theta_J$ for all system sizes. The dotted line shows the $\Theta_J^2$ scaling predicted when the effective viscosity ratio $\eta_{\rm eff}/\eta_f$ is approximately constant. The data follow this prediction over more than two decades for the larger systems, with deviations developing at high $\Theta_J$ for the smallest systems, where the local $J$ is largest.}
\label{fgr:theta}
\end{figure}

\begin{figure}[!htbp]
 \includegraphics[width=0.46\textwidth]{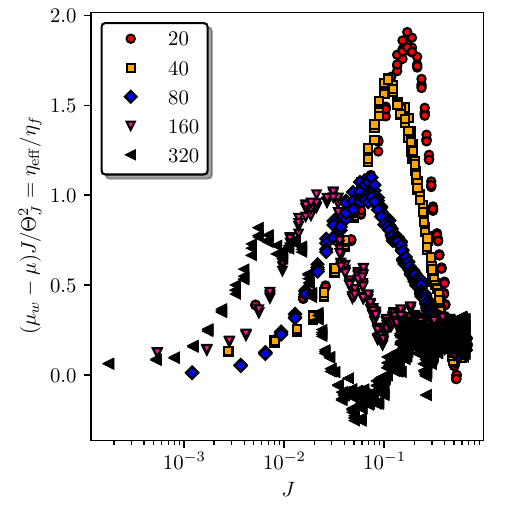}
 \caption{Extracting \(\eta_{\rm eff}/\eta_f\) as a function of \(J\) from Eq.~\ref{eq:covariance_theta}. 
 Non-monotonic behaviours are observed for all system sizes.
 At low \(J\), near the shear‑reversal regions, the effective viscosity exhibits a clear shear‑thickening behaviour, despite a concomitant decrease in the packing fraction (see, \emph{e.g.}, Fig.~\ref{fgr:phi_mzz}). After reaching a peak viscosity, the viscosity decreases as a function of $J$.
 The decrease at high $J$ is similar to a shear-thinning behaviour, but could also be driven by a decreasing packing fraction at higher $J$ values. Some bins yield slightly negative values, particularly for the largest system at intermediate $J$, where $\langle \delta \tau \delta \dot \gamma \rangle$ probably becomes small and statistical noise can drive the correlation through zero. All of the above indicates that the constitutive assumption $\delta \tau \approx \eta_{\rm eff} \delta \dot \gamma$ is not strictly $J$-independent.
 }
\label{fgr:etaeff}
\end{figure}

If, in addition, the \emph{fluctuating} response of the contact and lubrication network is approximately viscous, $\delta
\tau \approx \eta_{\mathrm{eff}}\,\delta\dot\gamma$, then
\begin{equation}
\frac{\langle \delta\tau\,\delta\dot\gamma \rangle}{\sigma_{yy}\langle \dot\gamma \rangle}
\approx \eta_{\mathrm{eff}}\, \frac{\langle \delta\dot\gamma^2 \rangle}{\sigma_{yy}\langle\dot\gamma \rangle}  
\propto \eta_{\mathrm{eff}}\,\frac{\sigma_{yy}}{\eta_f^2 \langle\dot\gamma \rangle} \,\Theta_J^2 =\frac{\Theta_J^2}{J} \frac{\eta_{\mathrm{eff}}}{\eta_f},
\label{eq:covariance_theta}
\end{equation}
and the correction in Eq.~(\ref{eq:tauW_decomposition}) scales as
$\Theta_J^2/J$, precisely the dimensionless combination that controls the cooperative correction in non-local models built on rescaled
granular temperature. The dissipative formulation does not contradict
these theories; rather, it absorbs the same physics directly into a
stress measure and bypasses $\Theta_J$ as an independent state
variable. This also explains why power-law collapses in
$(\mu,\Theta_J,J)$ succeed empirically while offering little
mechanistic insight: they effectively measure the covariance
$\langle \delta\tau\,\delta\dot\gamma \rangle$ through the proxy
$\langle \delta\dot\gamma^2 \rangle$, under the implicit assumption
that the fluctuating response is viscous.
This perspective also makes the constitutive assumption underlying both approaches explicit. When $\delta\tau$ and $\delta\dot\gamma$ are
tightly correlated through an effective viscosity, as appears to be the case for frictionless dense suspensions, $\tau_W$ collapses
inhomogeneous data onto the homogeneous $\mu(J)$ law, as shown in Fig.~\ref{fgr:MuJ}. 
Figure \ref{fgr:theta} tests this connection directly. Plotting $(\mu_W-\mu)J$ against $\Theta_J$ across all system sizes shows that the data collapse onto an approximate $\Theta_J^2$ scaling over more than two decades, in agreement with Eq.~\ref{eq:covariance_theta} under the assumption that $\eta_{\rm eff}/\eta_f$ is approximately constant. Systematic deviations emerge for the smallest systems, signalling that $\eta_{\rm eff}/\eta_f$ varies with the local $J$, $\Theta_J$, $\mu$, and $\mu_W$ values more than in the larger (close to bulk-like) systems. 
To test this hypothesis, we extract $\eta_{\rm eff}/\eta_f$ directly from Eq.~\ref{eq:covariance_theta} (Fig.~\ref{fgr:etaeff}). The effective viscosity is clearly non-monotonic. At low $J$, near the shear reversals, $\eta_{\rm eff}/\eta_f$ rises sharply with $J$, despite the concomitant decrease in packing fraction (Fig.~\ref{fgr:phi_mzz}), and the peak height shrinks systematically as the system size $L_y/\langle d\rangle$ increases. Beyond the peak, $\eta_{\rm eff}/\eta_f$ decreases with $J$ in a manner reminiscent of shear thinning, although this trend is consistent with, and may be primarily driven by, the falling packing fraction at higher $J$. Together, these observations indicate that the constitutive assumption 
$\delta \tau \approx \eta_{\rm eff} \delta \dot \gamma$ underlying Eq.~\ref{eq:covariance_theta} is not strictly $J$-independent, although it remains a useful first-order approximation in the small-$J$ limit relevant to apparent non-locality.



\subsection{Statistical convergence near reversals}
\begin{figure}[!htbp]
 \includegraphics[width=0.235\textwidth]{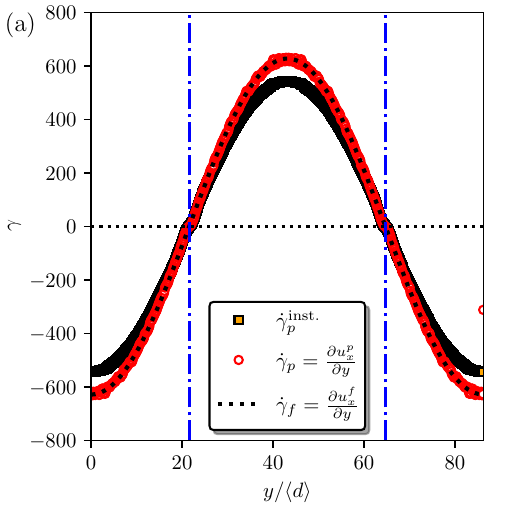}
  \includegraphics[width=0.235\textwidth]{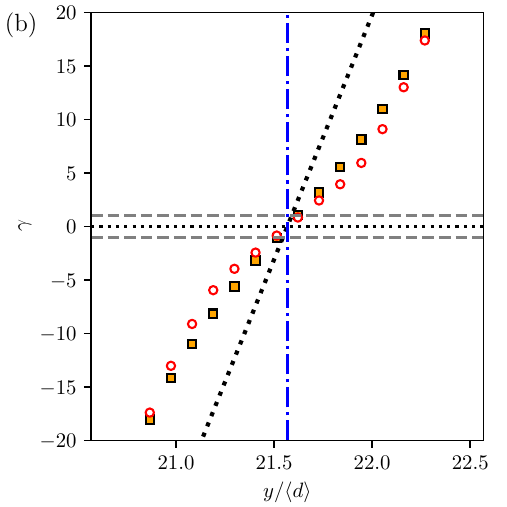}
 \caption{Typical accumulated strain in the collected data for $L_y/\langle d\rangle \simeq 80$: (a) entire system and (b) region of interest near the shear reversal.}
\label{fgr:strain}
\end{figure}

Because the strain rate varies spatially, the accumulated strain is nonuniform across the system, and data from different regions correspond to different strain histories. The region of primary interest, near the shear reversals, experiences the smallest accumulated strain; nevertheless, all data presented in the manuscript correspond to accumulated strains of order unity, and often to larger values. Figure~\ref{fgr:strain} illustrates the spatial variation of accumulated strain in the collected data with the box length $L_y/\langle d \rangle \simeq 80$.

\end{document}